# Imaging of Microscopic Sources of Resistive and Reactive Nonlinearities in Superconducting Microwave Devices

A. P. Zhuravel, Steven M. Anlage, *Member, IEEE,* and A. V. Ustinov

*Abstract*— The technique of low-temperature Laser Scanning Microscopy (LSM) has been applied to the investigation of local microwave properties in operating $YBa_2Cu_3O_7/LaAlO_3$ thin-film resonators patterned into a meandering strip transmission line. By using a modified newly developed procedure of spatially-resolved complex impedance partition, the influence of inhomogeneous current flow on the formation of nonlinear (NL) microwave response in such planar devices is analyzed in terms of the independent impact from resistive and inductive components. The modified procedure developed here is dramatically faster than our previous method. The LSM capability to probe the spatial variations of two-tone, third-order intermodulation currents on micron length scales is used to find the 2D distribution of the local sources of microwave NL. The results show that the dominant sources of microwave NL are strongly localized in the resistive domains.

*Index Terms*—Laser scanning microscopy, microwave devices, intermodulation, nonlinearity, high-$T_c$ superconductors.

## I. Introduction

IDENTIFICATION of *microscopic* sources of resistive and reactive nonlinearities (NLs) in high-temperature superconducting (HTS) films, as well as clarification of the distinct role that such sources are playing in the *macroscopic* (integral) response of passive microwave devices, is a vital issue of fundamental and applied research for HTS materials in high-frequency (HF) fields. For example, the examination of the nonlinear Meissner effect (NLME) for evidence of d-wave gap symmetry in the cuprates [1] is facilitated by inductive NL at current-carrying strip edges. In turn, the HF current-dependent topology of the *microscopic* sources of resistive NLs is significant for explanation of NL response of a HTS device in the critical state [2]. And finally, improvement of microwave device technology comes from a thorough understanding of the contribution of different defects, the topology of dc and rf transport, as well as superconducting, geometrical, and structural inhomogeneities.

We have shown earlier that the low-temperature Laser Scanning Microscope (LSM) is a powerful nondestructive evaluation technique for non-contact, spatially-resolved probing of the local HF current density, $J_{HF}(x,y)$ in HTS films and devices [3]. This LSM method was applied next for mapping optical, thermal, HF and dc electron transport properties and superconducting critical parameters of HTS samples directly in their operating state at $T < T_c$, with micron-scale spatial resolution [4]. Furthermore, a procedure of spatially-resolved partition of its resistive and inductive photo-response (PR) components is developed in [5], while a procedure of LSM PR(x,y) calibration to determine the absolute amplitude of $J_{HF}(x,y)$ is described in [4].

In this paper, we develop a new procedure of rapid, spatially-resolved LSM analysis of HTS devices in varied HF fields to investigate the 2D distribution of the local sources of microwave NL in a microstrip-line resonator. Our results show that a spatial topology of dominant NL sources is correlated with the microwave-field-dependent distribution of the resistive domains formed by regions of local overcritical current densities in the superconducting strip.

## II. Experimental Details

### A. Sample

The sample is a $YBa_2Cu_3O_y$ (YBCO) film with thickness of about 1 μm configured by ion-milling lithography on $LaAlO_3$ (LAO) substrate into a meandering strip line resonator with line width of 250 μm. The geometry of the resonator is shown in Fig.1. The frequency of fundamental resonance is about 1.85 GHz. Here, we give an example of LSM characterization of the resonator at the third harmonic frequency of about 5.96 GHz where it demonstrates a loaded $Q_L \sim 2000$ at T=80 K. The device is capacitively coupled to an input RF circuit delivering power $P_{IN}$ in the range from –40 to +10 dBm and mounted in a copper microwave package. The package was cooled inside the vacuum cavity of an optical cryostat which stabilizes the temperature of the sample in the range *77-95* K with an accuracy better then 5 mK [3,4].

Manuscript received August 28, 2006. This work is supported in part by NASU grant "Nanosystems, nanomaterials and nanotechnologies", the German Science Foundation (DFG), and by NSF/GOALI DMR-0201261, NSF/DMR-0302596, and NSF/ECS-0322844.

Alexander P. Zhuravel is with B. Verkin Institute for Low Temperature Physics & Engineering, National Academy of Sciences of Ukraine, 61164 Kharkov, Ukraine (phone: +38 (057)341-0907; fax: +38(057)345-0593; e-mail: zhuravel@ilt.kharkov.ua).

Steven M. Anlage is with the Physics Department, Center for Superconductivity Research, University of Maryland, College Park, MD 20742-4111 USA.

Alexey V. Ustinov is with Physics Institute III, University of Erlangen-Nuremberg, D-91058, Erlangen, Germany (e-mail: ustinov@physik.uni-erlangen.de).

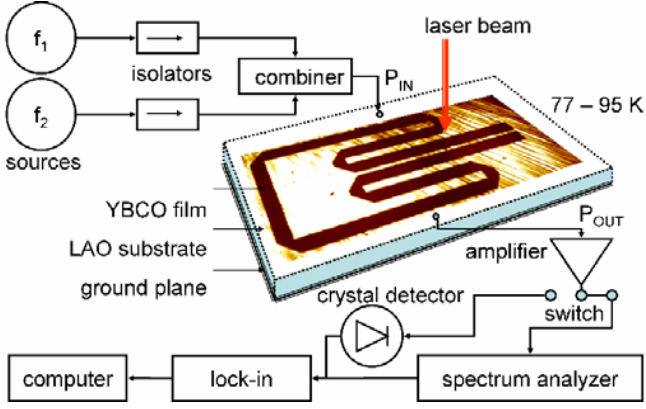

Fig. 1. Schematic of the microwave LSM experiment.

### B. Imaging procedures

The LSM studies were performed by *x-y* scanning of the resonator surface by a 1.1 μm diameter laser beam. The power of the laser $P_L$=10μW was fixed low enough to introduce minimal perturbation on the global microwave properties of the resonator. The intensity of the laser is modulated at a frequency of $f_M$=100 kHz, producing an oscillating thermal probe of about 4 μm in diameter [3,4,6,7]. Fig.1 shows a schematic of the experiment. Two fixed frequency signals ($f_1$ and $f_2$) are applied to the HTS resonator. These signals are centered on the $|S_{21}(f)|$ curve with an equal spacing ±Δf/2=500 kHz around the device resonant frequency $f_0$ and have the same amplitude $P_{IN}(f_1)=P_{IN}(f_2)$.

Transmitted HF signals are amplified and measured by a spectrum analyzer, as shown in Fig.2. In addition to $P_{f1}$ and $P_{f2}$ output tones, the spectrum of the resonator contains intermodulation harmonics $-IMD_3=P_{2f1-f2}$ and $+IMD_3=P_{2f2-f1}$ resulting from NL mixing of the signals at $f_1$ and $f_2$, as well as laser modulation sidebands. A zero-frequency span mode of the spectrum analyzer combined with a lock-in amplifier technique was used to separate the probe-induced modulation of the output spectrum (so-called LSM photoresponse, PR) at different *f* specified by the filled circles in Fig.2. Changes in $P_{f1}$ and $P_{f2}$ as a function of position *(x,y)* of the laser beam perturbation on the sample were collected to build up 2D maps of both resistive and inductive components of LSM PR in compliance with the partition method described in [5]. The intermodulation (IMD) imaging LSM mode was applied to image the distribution of *microscopic* sources of NL response at ±$IMD_3$ [6].

### III. RESULTS AND DISCUSSION

In our previous papers [6,7] examining microwave resonators of complex meander-line geometry we have demonstrated that the local sources of microwave NLs are non uniformly distributed across the film, and that IMD PR is dominantly localized in only small regions of the HTS film having the highest HF current densities, $J_{HF}(x,y)$. These regions coincide with the inner corners of the resonator structure and can radically affect the linearity of the device performance.

Fig.3(a) shows the optical reflectance LSM image collected in a 50 μm x 50 μm area of maximum $J_{HF}(x,y)$. The inductive $PR_X(x,y)$ image in this area is shown in Fig.3(b). The distribution of $PR_X(x,y)$ clearly shows the current bunching at the micro-strip edges and reaches two extrema corresponding to $J_{PEAK}$ ~ (3-4)x$10^9$ A/m$^2$ close to the inner corner at T = 78 K, and $P_{f1} = P_{f2}$ = -6 dBm. In addition, a spatial modulation of the $PR_X(x,y)$ is noticeable along the edges away from the corner. This may be caused by spatial variation of the magnetic penetration depth in different twin-domain blocks (TDB) of the HTS film formed as a result of twinning in the LAO substrate that is evident from Fig.3(a) [8]. As an example, arrow "A" in Fig.3 indicates the position of the interface between two TDB. A very small increase of the $PR_X(x,y)$ can be seen at this location.

In contrast, the resistive (see Fig.3(c)) component, $PR_R(x,y)$, demonstrates a sharper distribution of the LSM PR. It is also attached to the areas of local $J_{HF}(x,y)$ peaks at the inner corner and at the TDB interfaces. The origin of resistive response at that TDB interface (see point "A") may be connected with easier HF vortex entry in the HTS strip through suppression of the edge (Bean–Livingston) barrier there.

Fig.3(d) shows the 2D variation of NL (or IMD) LSM PR acquired simultaneously. With the exception of some details, there is spatial correlation between both sources of $PR_{IMD}(x,y)$ and $PR_R(x,y)$ visible in the images suggesting a significant resistive origin of the sources of NL response. Therefore, we studied the HF power $P_{IN}$-dependent spatial evolution of the $PR_R(x,y)$ to see how the NL sources originate.

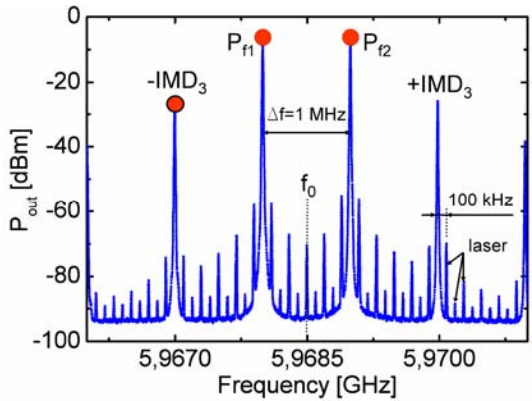

Fig. 2. Spectrum of output signal for $P_{IN}(f_1) = P_{IN}(f_2)$ = +12 dBm, T = 78 K, and $f_1$ = 5.968 GHz, $f_2$ = 5.969 GHz, and $f_M$ = 100 kHz.

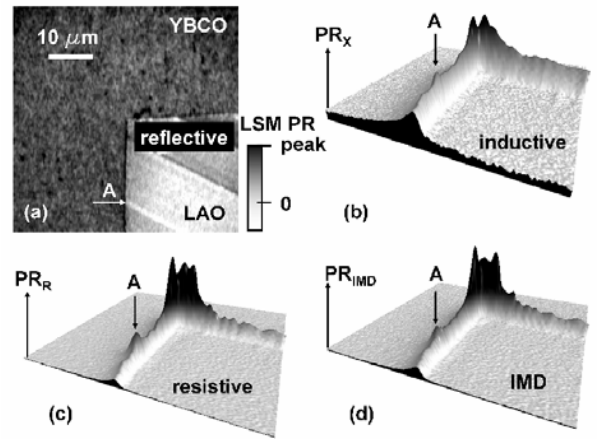

Fig. 3. 2-D LTLSM images of (a) reflectivity map showing the twinned structure of the substrate along with distribution of (b) inductive, (c) resistive, and (d) IMD components of LSM PR that are obtained in the same 50 μm x 50 μm$^2$ area around the inner corners of the resonator

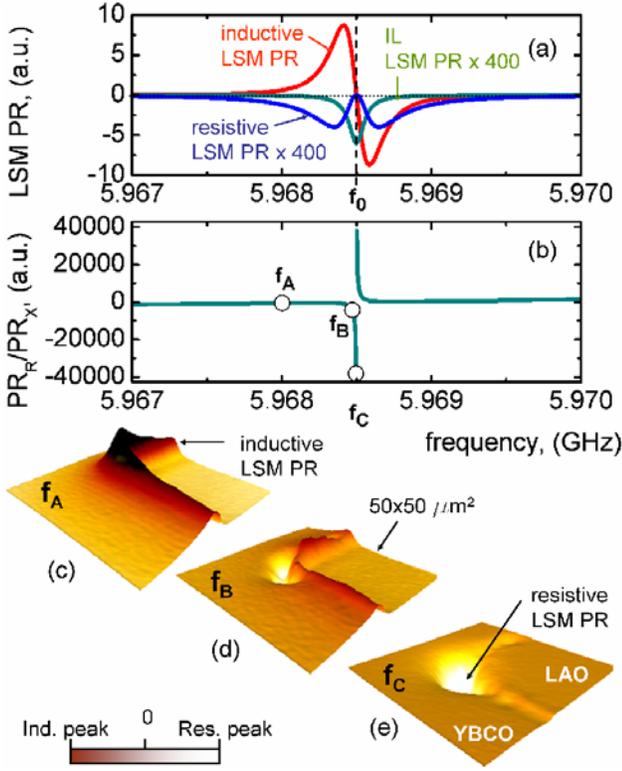

Fig. 4. Frequency dependence of (a) resistive, inductive and insertion loss (IL) components of LSM PR, (b) ratio of the sum of resistive and IL to inductive components, and 2D LSM images showing (c) image of mainly inductive LSM PR at $f_A$ far from $f_0$, (d) image with comparable amplitudes of resistive and inductive LSM PR components at $f_B$ closer to $f_0$ and (e) mainly resistive contribution at $f_C$ almost equal to $f_0$.

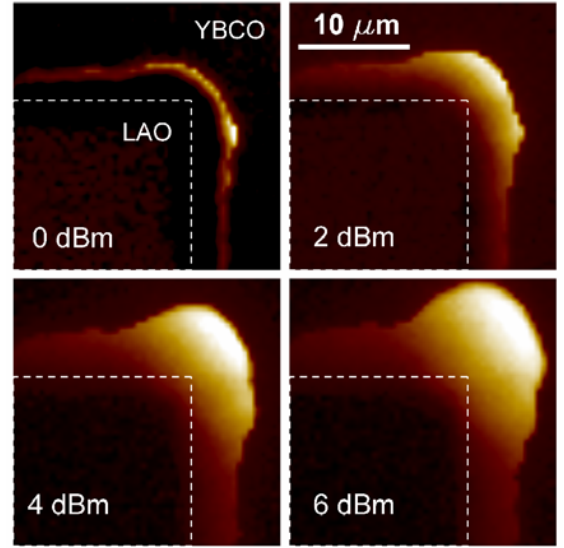

Fig. 5. Images of resistive LSM PR penetrating into the inside corner of an HTS film at the different input HF powers indicated in the images. White dotted boxes show the YBCO/LAO patterned edge. Brighter regions correspond to larger amplitude of $PR_R(x,y)$.

It is important to keep in mind that the total LSM PR originates from two independent (resistive and inductive) contributions. The $PR_R(x,y)$ component is directly related to photo-induced modulation of the inverse $Q$-factor of the resonator due to an increase in local Ohmic dissipation. On the other hand, the $PR_X(x,y)$ component is associated with the change of kinetic inductance leading to the effects of HF resonant frequency tuning [5]. For high-quality resonant devices (with Q>1000) the inductive component of LSM PR dominates, and so we did not see any visible power-dependent redistribution of LSM PR (x,y) at frequencies $f_1$ and $f_2$. Under such conditions, it is very labor-intensive to work with the set of LSM images obtained at $f_1(P_{IN})$ and $f_2(P_{IN})$ to construct the $PR_R(x, y, P_{HF})$ using the method described in [5].

Here we propose a new procedure of rapid, spatially-resolved approximate identification of the $PR_R(x,y)$ component using a simpler experimental setup employing the crystal detector instead of the spectrum analyzer in Fig. 1. In this case only a single input tone is used. The procedure is based on the sharp increase of the ratio $PR_R/PR_X$ at a frequency close to $f_0$. Fig.4(a) shows the frequency dependence of $PR_R$ and $PR_X$, calculated from Eqs. (8) and (9) in [4] describing these responses in terms of laser-beam-induced changes of resistive and inductive components of LSM PR, respectively. This model was revised slightly to include the contribution of insertion loss (IL) contrast to LSM PR, as shown in Fig. 4(a). We made the approximation that the IL contribution is mainly resistive in nature and can be added to the total resistive LSM PR. With this assumption, the ratio $PR_R/PR_X$ is plotted in Fig. 4(b). It is seen that this ratio may exceed $10^4$ at $f_C \sim f_0$ compared to a value of $10^{-3}$ off resonance. Such an appreciable difference means that the LSM image obtained at $f_A$ reflects mainly inductive HF properties of the resonator while the LSM image obtained at $f_C$ shows the distribution of resistively-induced PR.

As an example, Fig.4(c) illustrates the distribution of LSM PR around an inside corner of the resonator at fixed frequency $f_A$ = 5.9675 GHz at T = 78 K, and $P_{fA}$ = -24 dBm. This distribution almost perfectly repeats the inductive PR that was extracted using the method of component partition [5]. Fig. 4(d) shows a spatial modification of the LSM PR(x,y) profiles for the case when both resistive and inductive components become comparable in their influence at $f_B$ = 5.9682 GHz close to $f_0$. In contrast, Fig. 4(e) obtained under the same conditions at $f_C$ = 5.9685 GHz manifests the appearance of resistive LSM PR whose amplitude is much higher than the inductive component at $P_{IN}(f_C)$ = -24 dBm. We establish that for input powers $P_{IN}(f_C)$ > -12 dBm it was impossible to observe a difference between the $PR_R(x,y)$ distributions measured by both two- and one-tone imaging procedures. Thus, the excitation of the resonator at a single frequency $f_C$ = 5.9685 GHz was used to image the spatial penetration of the resistive LSM PR into the HTS film at different input HF power.

Fig. 5 shows that the first resistive areas in the HTS film form around the inside corner on a curved line for $P_{IN}(f_C) \leq 0$ dBm inside the strip edge. By comparison with the $PR_X(x,y)$ distribution (not shown) in this 25x25 μm$^2$ area it was determined that the position of the resistive line coincides with the regions of the HTS film carrying the highest $J_{HF}(x,y)$.

With increasing $P_{IN}(f_C)$ from +2, +4 and then +6 dBm, the $PR_R(x,y)$ occupies increasingly more area of the HTS film surrounding the inner corner of the resonator structure. It is seen that the $PR_R(x,y)$ contrast has a very sharp front inside the HTS film and a smooth decay toward the patterned edges. The same resistive patterns were observed at all the corners of this resonator and in other HTS structures of meandering geometry imaged earlier. This is reminiscent of the development of an rf critical state in the film.

Such behavior is similar to that observed as a large disturbance of the flux pattern produced by increasing the applied magnetic field around macroscopic individual defects near the strip edge [9]. Here the role of the defect is played by the nearly zero-radius corners creating very large values of $J_{HF}(x,y)$ as a result of the sharp curvature of HF current flow. However, here we can not confirm the magnetic origin of the resistive domains due to the fact that the inductive $PR_X(x,y)$ does not change its shape at any $P_{IN}$ applied up to +12 dBm. Also, we did not detect clear trajectories of magnetic penetration or the influence of TDBs on the formation of the resistive pattern. One possible explanation is current-induced resistive dissipation in zones of preferred HF vortex penetration and hysteresis [10]. A definitive answer may be established by combining these results with measurements of the same resonator in other LSM dc/HF imaging modes [4].

To show the difference between (i) HF current and (ii) magnetic-field-induced sources of resistive NL, we studied the spatial and power-dependent evolution of $J_{HF}(x,y)$ and resistive zones in the HTS resonator in the presence and absence of external static magnetic field. Fig. 6(a) shows a 125x25 μm² map of LSM PR of the zero-field-cooled (to 78 K) resonator obtained at $f_1 = 5.968$ GHz in the area including a natural grain boundary (GB) marked by the horizontal white dashed line. This GB starts from the strip edge (black dashed line) and crosses the interfaces of TDBs (white dashed bevels) through the whole HTS structure. Only the inductive component (seen as a bright elongated spot along the GB) of LSM PR(x,y) was detected in this environment in the range of applied HF power from -40 dBm to +10 dBm. The shape of this pattern was practically independent of $P_{IN}$ and has the typical $J_{HF}(x,y)$ peak near the edge and characteristic spatial modulation along TDBs [8]. Fig.6(b) shows the distribution of LSM PR modified at $P_{IN} = +10$ dBm by a c-oriented, DC magnetic field of strength 2-10 G from a permanent magnet. A radical transformation of the LSM PR(x,y) map is evident. The inductive pattern looks brighter due to the DC magnetic field suppressing the critical current in the GB. Dark resistive areas have been created by introduction of vortices at the intersection of the GB and TDBs shown by black arrows. The size of the resistive domains is equal to the size of the laser probe showing that the vortices are strongly pinned by the defects of the HTS structure. This feature clearly shows the difference between dc magnetically-derived resistivity, which is frequently used to explain NL properties of HTS films, and the HF current-induced losses like those presented in Fig. 5.

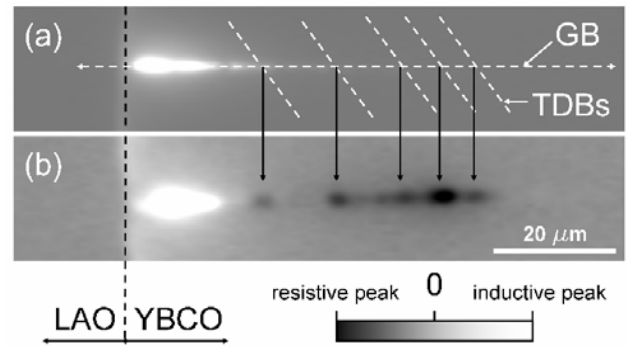

Fig. 6. 2D patterns of LSM PR obtained at 78 K, +10 dBm in a 125x25 μm² area of the HTS resonator in the (a) absence and (b) presence of DC magnetic field. Bright (dark) spots correspond to inductive (resistive) response while grey regions are of zero PR signal. The edge of the strip is indicated by the black dashed line. Positions of the GB and TDBs coincide with the white dashed lines.

ACKNOWLEDGMENT

We acknowledge Stephen Remillard (Agile Devices, USA) for providing the high-quality HTS samples.